# Antiferrodistortive Phase Transition in Pseudorhombohedral $(Pb_{0.94}Sr_{0.06})(Zr_{0.550}Ti_{0.450})O_3$ : A Combined Synchrotron X-ray and Neutron Powder Diffraction Study

Ravindra Singh Solanki, [1] S. K. Mishra, [2] Anatoliy Senyshyn, [3] I. Ishii, [4] Chikako Moriyoshi, [5] Takashi Suzuki, [4] Yoshihiro Kuroiwa[5] and Dhananjai Pandey[1]

[1]School of Materials Science and Technology, Indian Institute of Technology (Banaras Hindu University), Varanasi-221005, India

[2]Research and Technology Development Centre, Sharda University, Greater Noida-201306, India

[3]Forschungsneutronenquelle Heinz Maier-Leibnitz (FRM II), Technische Universit¨at M¨unchen, Lichtenbergstrasse 1,D-85747 Garching bei M¨unchen, Germany

[4]Department of Quantum Matter, ADSM, Hiroshima University, Japan

[5]Department of Physical Science, Graduate School of Science, Hiroshima University, Japan

## Abstract

The controversies about the structure of the true ground state of pseudorhombohedral compositions of $Pb(Zr_xTi_{1-x})O_3$ (PZT) are addressed using a 6% $Sr^{2+}$ substituted sample with x=0.550. Sound velocity measurements reveal a phase transition at $T_c$~279K. The temperature dependence of FWHM of $(h00)_{pc}$ peaks and the unit cell volume also show anomalies around 279K even though there is no indication of any change of space group in the synchrotron X-ray powder diffraction (SXRD) patterns. The neutron powder diffraction patterns reveal appearance of superlattice peaks below $T_c$~279K confirming the existence of an antiferrodistortive phase transition. The Rietveld analysis of the room temperature and low temperature SXRD data below $T_c$ shows that the structure corresponds to single monoclinic phase in the Cm space group while the analysis of neutron powder diffraction data reveals that the structure of the ground state phase below $T_c$ corresponds to the Cc space group. Our analysis shows that the structural models for the ground state phase based on R3c space group with or without the coexistence of the room temperature monoclinic phase in the Cm space group can be rejected.

## I. Introduction

Lead Zirconate Titanate ($Pb(Zr_xTi_{1-x})O_3$), commonly known as PZT, is the most widely used piezoelectric ceramic in sensor and actuator devices[1]. It displays a rich and complex x-T phase diagram which contains a nearly vertical morphotropic phase boundary (MPB) at x=0.520, a composition for which the piezoelectric and dielectric responses are highest[1]. The MPB has traditionally been believed to separate the stability fields of rhombohedral (x>0.520) and tetragonal (x<0.520) phases in the R3m and P4mm space groups, respectively[1]. In addition to the MPB, PZT shows two more phase boundaries on the Zr-rich side of the phase diagram. One such boundary occurs at $x \gtrsim 0.940$ and separates the stability field of an orthorhombic phase (space group Pbam) stable for $x \gtrsim 0.940$ from that of a second rhombohedral phase in the R3c space group. *In this R3c phase,* the oxygen octahedra are tilted about the trigonal axis[2] of the neighbouring unit cells in an antiphase manner and therefore the structure belongs to the $a^-a^-a^-$ tilt system[3]. This rhombohedral phase with antiphase tilted octahedra results from a high temperature rhombohedral phase with untilted octahedra in the R3m space group through an antiferrodistrotive (AFD) phase transition[1]. The AFD phase transition temperature first increases with decreasing Zr content, then peaks at $x \approx 0.850$ and finally starts decreasing until for $x \approx 0.620$ the room temperature phase corresponds to the R3m space group with untilted octahedra. This gives rise to another phase boundary at $x \approx 0.620$ separating the stability fields of the two different rhombohedral phases. All the phases (tetragonal P4mm, rhombohedral R3m, rhombohedral R3c and orthorhombic Pbam) transform to the cubic paraelectric phase (Pm3m) and the corresponding transition temperature decreases with increasing $Zr^{4+}$ content, irrespective of the structure of the room temperature phase.

While the AFD phase transition in PZT has been well known ever since the phase boundary above room temperature between the 'R3m' and 'R3c' rhombohedral phases was established[1], it was to the credit of Ragini et al[4] and Ranjan et al[5] who discovered this transition below room temperature for compositions in the MPB region also, in agreement with a theoretical prediction by Fornari et al[6]. This discovery of AFD transition below room temperature was subsequently confirmed by several workers[7-13]. The phase below the AFD transition temperature is a superlattice phase of the higher temperature phase[4, 5]. The contentious issue under intense debate at present is ''what is the space group of the superlattice phases of PZT which represent the true ground state and result from a ferroelectric phase by an AFD transition either below room temperature or above the room temperature depending on the $Zr^{4+}$ content?''. To seek answer to this contentious issue, it is necessary to understand the subtle differences between three different types of monoclinic phases that exist or may exist at room temperature: (i) Firstly, there is a pseudotetragonal monoclinic phase with Cm space group for which the $(200)_{pc}$ peak is a doublet. As shown by Singh et al[14] and Pandey et al[15] using Rietveld analysis of high resolution synchrotron X-ray powder diffraction (SXRD) data, this phase exists over a very narrow composition range $0.520 \lesssim x < 0.530$ at room temperature, as confirmed by refinements. This is the phase discovered by Noheda et al[16, 17] to which the tetragonal phase transforms below room temperature for x=0.520. This phase exists in the vicinity of the MPB only and its stability field gets narrower with increasing temperature and undergoes an AFD transition to the superlattice phase below room temperature[14, 15].

The space group of this monoclinic phase was first shown by Hatch et al[18] to be Cc which was subsequently confirmed by others[11-13] but an alternative model based on the coexistence of R3c+Cm phases has also been proposed by some workers[8,9]. (ii) Secondly, there is a pseudorhombohedral monoclinic phase also in the Cm space group for which the $(200)_{pc}$ is a singlet. The characteristic feature of this phase is anomalously large broadening of $(h00)_{pc}$ and $(hh0)_{pc}$ reflections as was first pointed out by Ragini et al[19] and subsequently confirmed by Singh et al[14] in their high resolution SXRD studies. Following Ragini et al's[19] work, other workers have also accepted the assignment of Cm space group to such 'pseudorhombohedral' compositions[20-24]. The anomalous broadening of the $(h00)_{pc}$ and $(hh0)_{pc}$ reflections of the pseudorhombohedral monoclinic phase disappears in the tetragonal phase stable at $x<0.520$[19]. This phase has been traditionally assigned the R3m space group and is stable in the composition range $0.530 \leq x \leq 0.620$ in chemically homogeneous samples[14,15,19]. (iii)Thirdly, there is a rhombohedral phase[1] or a pseudorhombohedral monoclinic phase[14, 15] for $0.620 \lesssim x \lesssim 0.940$ which also exhibits anomalous broadening of the $(h00)_{pc}$ reflections but to a lesser extent as compared to the case for $0.530 \leq x \leq 0.620$. The powder XRD pattern of this rhombohedral or pseudorhombohedral phase contains superlattice peaks at room temperature due to the cell doubling causes by antiphase tilting of oxygen octahedra in the structure. Traditionally, this phase has been assigned the rhombohedral R3c space group[1] but according to Pandey et al[15], this phase may also be monoclinic in the Cc space group. The AFD transition temperatures for the first two types of monoclinic phases occur below room temperature[15] and decreases with decreasing Zr content, whereas for the third type it occurs above room temperature[1]. This suggests the weakening of the driving force for the AFD transition on approaching the MPB from the Zr-rich side. The AFD transition eventually disappears for tetragonal compositions away from the MPB.

To settle the Cc versus R3c space group controversy for the ground state of pseudotetragonal monoclinic compositions of PZT, we recently[25] adopted a novel strategy combining the use of 6% $Sr^{2+}$ substitution at $Pb^{2+}$ site and high wavelength neutrons (2.44Å) to increase the tilt angle (and hence the intensity of the superlattice peaks) and separate the most intense lower angle superlattice peak at q= $(3/2\ 1/2\ 1/2)_{pc}$ from the neighbouring intense $(111)_{pc}$ perovskite peak, respectively. Using these two strategies, we could not only resolve the intense $(3/2\ 1/2\ 1/2)_{pc}$ superlattice peak from the tail of the neighbouring $(111)_{pc}$ peak but could also show that the $(3/2\ 1/2\ 1/2)_{pc}$ superlattice is not a singlet[25]. For the R3c space group, this peak has to be a singlet. It was thus concluded that the space group of the pseudotetragonal composition of $(Pb_{0.94}Sr_{0.06})(Zr_xTi_{1-x})O_3$ (PSZT) for x=0.530 is not R3c but Cc[25]. As a next step, we consider in the present work AFD transition in a pseudorhombohedral monoclinic composition $(Pb_{0.94}Sr_{0.06})(Zr_{0.550}Ti_{0.450})O_3$ (PSZT550) with a view to settle the R3c versus Cc space group controversy for such compositions[15, 20-24]. We present results of synchrotron X-ray diffraction (SXRD), neutron powder diffraction (NPD) and elastic modulus measurements which suggest that the R3c space group can be rejected for the pseudorhombohedral compositions as well. We also compare the Rietveld refinement results for R3c (including its coexistence with Cm space group[25,15]) and Cc space group models for both the pseudotetragonal (x=0.530 i.e., PSZT530) and pseudorhombohedral (x=0.550 i.e., PSZT550) monoclinic compositions to settle the existing controversies about the space group of the ground state of PZT in and around the MPB region.

## II. Experimental

Our samples were prepared by a semiwet route[26] which is known to give the narrowest composition width $\Delta x \approx 0.01$ of the MPB region for pure PZT. Full description of the sample preparation has been given in our previous work[25]. High resolution Synchrotron X-ray powder diffraction (SXRD) measurements were carried out in the 100 to 800K temperature range at the BL02B2 beam line of SPring-8, Japan[27] at a wavelength of 0.412Å (30 keV). Neutron powder diffraction (NPD) data has been obtained from FRMII, Germany on high resolution SPODI[28] powder diffractometer at a wavelength of 2.536Å. The longitudinal elastic modulus ($C_L$) has been obtained by measuring the sound velocity (v) using the phase comparison-type pulse echo method[29]. Rietveld refinements using Synchrotron and Neutron powder diffraction data were carried out using the FULLPROF software package[30].

## III. Results and discussion

### (A) Room Temperature phase

Fig. 1 depicts the SXRD profiles of the $(111)_{pc}$, $(200)_{pc}$ and $(220)_{pc}$ reflections at T≤300K. For the ease of discussion, we have also included in the figure, the SXRD profiles of these reflections in the cubic phase at 800K. For all the temperatures with T≤300K, the $(111)_{pc}$ and $(220)_{pc}$ are doublet, whereas $(200)_{pc}$ is a singlet, as expected for a rhombohedral structure. However, the $(200)_{pc}$ peak shows an anomalous broadening. The width of the $(200)_{pc}$ peak at T=300K is ~2.66 times that of the neighbouring $(111)_{pc}$ peak. A similar anomalous broadening of $(h00)_{pc}$ peaks was reported by Ragini et al[19] and Singh et al[14] in the so-called 'rhombohedral' compositions of PZT and Singh and Pandey[31] in PMN-xPT ceramics. The anomalous broadening of $(h00)_{pc}$ type reflections can be attributed to anisotropic strains which may result from the size difference between the $Zr^{4+}$ and $Ti^{4+}$ ions at the B-site and/or $Sr^{2+}/Pb^{2+}$ ions at the A-site of the $ABO_3$ perovskite structure. It can also result from local compositional fluctuations. Anisotropic strain broadening can be modeled using Stephens's model[32] in the Rietveld refinements assuming the rhombohedral R3m space group. Unlike Frantti et al[8] for x=0.530 at room temperature and Zhang et al[21] for x=0.60 at 473K, who considered R3c space group for pseudorhombohedral compositions, we did not find any necessity of considering the R3c space group as the superlattice reflections resulting from $a^-a^-a^-$ oxygen octahedral tilt expected for this space group are not observed in the SXRD patterns. The intense superlattice peak $(3/2\ 1/2\ 1/2)_{pc}$ for the R3c space group is expected to occur at d≈2.452Å in the SXRD data but no such peak is observed as can be seen from Fig. 1(a) of the supplementary file[33]. The intensity of the $(3/2\ 1/2\ 1/2)_{pc}$ peak depends on small changes in the oxygen positions in the R3c space group with respect to those in the R3m space group. Since X-rays cannot accurately detect such small changes in the positions of low 'Z' elements like oxygen, wherea*s* the neutrons can locate the positions of oxygen atoms more accurately, we searched for the $(3/2\ 1/2\ 1/2)_{pc}$ superlattice peak in the neutron powder diffraction pattern also recorded at 300K (see Fig. 1(b) of the supplementary file[33]). But we do not find any evidence for this peak at room temperature even in the neutron diffraction pattern. Accordingly, the R3c space group model for PSZT550[8,21] can be rejected at room temperature leaving behind R3m space group as a plausible structural model.

We therefore refined the structure using R3m space group considering all permitted anisotropic strain broadening parameters. The results are shown in Fig. 2(a) which reveals misfits for all $(hhh)_{pc}$, $(h00)_{pc}$ and $(hh0)_{pc}$ reflections as can be seen from the inset. The value of $\chi^2$ is 15.5 which is very high for a refinement to be crystallographically acceptable. It may therefore be concluded that the R3m space group cannot account for all the peaks consistently. On the other hand, consideration of the monoclinic space group Cm along with anisotropic strain broadening gave excellent fit between the observed and calculated profiles as can be seen from Fig.2 (b). The $\chi^2$ dropped from a value of 15.5 for the best refinement using R3m space group to 2.95 for the Cm space group model even though anisotropic strain broadening was considered in both the cases. Yokota et al[20] and Zhang et al[21] for similar 'pseudorhombohedral' compositions have found that pure Cm space group cannot account for the observed diffraction profiles and they had to include a coexisting phase with R3m/R3c space group in their refinements to obtain satisfactory Rietveld fits. Accordingly, we considered coexistence of two phases with Cm and R3m space groups also in our refinements but the coexisting R3m phase was rejected in the refinement giving nearly zero phase fraction for this phase. This confirmed that our sample of PSZT550 is single phase monoclinic in the Cm space group at 300K. The refined parameters for the Cm space group are given in the Table 1 of the supplementary files[33]. We believe that the coexisting phases reported by other workers[20-23, 34] may essentially be the artifacts of chemical heterogeneities in the sample and are caused by the difference in the sample preparation technique[35].

**(B) Low Temperature Structural Studies using SXRD**

It is evident from Fig. 1 that there is no evidence for any additional peak splitting down to 100K. Frantti et al[8] and Yokota et al[20] have proposed that the room temperature monoclinic phase in the Cm space group undergoes an AFD transition to the rhombohedral phase in the R3c space group, even though the parent monoclinic phase continues to coexist with the rhombohedral phase at low temperatures. Since the superlattice reflections of the R3c space group are not discernible in the SXRD profiles of PSZT550 even at 100K (see Fig. 2(a) of the supplementary files[33]), the Cm to R3c transition would appear as Cm to R3m transition in the SXRD patterns. Accordingly, we refined the structure at 100K using the SXRD data for the R3m, R3m+Cm and Cm models. In all the refinements, anisotropic strain broadening was used. Fig. 3(a) depicts the fits using the R3m model. It is evident from this figure that the R3m model does not explain the observed profiles as it leads to huge misfits and abnormally large $\chi^2$ values, similar to the situation at 300K. The consideration of Cm model not only led to excellent fit but also to a substantial reduction in the value of $\chi^2$ from 20.4 (the best value that could be obtained for the R3m space group) to 2.91 at 100K (see Fig. 3(b) for the Rietveld fit). Attempts to carry out two phase refinement using R3m+Cm model of Frantti et al[8] and Yokota et al[20] was reduced to Cm phase model giving zero phase fraction for the coexisting R3m phase, as was the case at 300K. Thus, SXRD data do not reveal any evidence for Cm to R3c/R3m or Cm to R3c/R3m+Cm transition. We can therefore reject the proposition of Cm to R3m/R3c transition given in Ref. 8 and 20. Table 1 of the supplementary file[33] compares the refined parameters at 100K with those at 300K.

**(C) Indication of a phase transition at low temperatures**

The Cm to R3m/ R3c+Cm or Cm to R3m/R3c transition proposed in Ref. 8 and 20 is expected to be a first order transition since R3m is not a subgroup of the Cm space group. Accordingly, such a transition should lead to a small discontinuous change in the unit cell volume. The variation of unit cell volume of the monoclinic Cm phase, obtained by Rietveld refinement at various temperatures, is depicted in Fig.4 (a). This figure does not reveal any discontinuous change in the unit cell volume suggesting the absence of any first order transition below room temperature. However, there is a gradual change of slope around 270K in Fig. 4(a). The linear interpolations from both the high and low temperature sides intersect at~275K suggesting that there may be a second order phase transition around this temperature.

As pointed out earlier, since the FWHM of $(200)_{pc}$ peak is ~2.66 times that of the neighbouring $(111)_{pc}$ peak, the structure could not be refined using the R3m space group. If there is a Cm to R3m phase transition below room temperature, one expects a decrease in the FWHM of the $(h00)_{pc}$ peaks. However the FWHM of the $(h00)_{pc}$ peaks is found to increase with decreasing temperature as shown in Fig. 4(b) for the $(200)_{pc}$ peak. Such an increase does not favour rhombohedral R3m space group. It is interesting to note that the FWHM increases nearly linearly from 400K down to ~275K beyond which it shows a quadratic dependence on temperature indicating a phase transition around 275K.

The existence of a phase transition at ~275 K was confirmed by the observation of an anomaly in the temperature dependence of the elastic modulus of PSZT550 around 275K. Fig. 4(c) depicts the longitudinal elastic modulus ($C_L$) of PSZT550, obtained from ultrasonic velocity measurements as a function of temperature. The longitudinal elastic modulus decreases with decreasing temperature up to 279K. Below 279K, the longitudinal elastic modulus starts increasing. A decreasing value of the elastic modulus with decreasing temperature is anomalous and is a signature of lattice instability[25]. The fact that below 279K the temperature dependence of $C_L$ assumes normal behavior reveals a phase transition at $T_c$~279K. The inset of Fig. 4(c) shows a magnified view of the transition region measured during heating and cooling cycles. The value of thermal hysteresis in $T_c$ is less than 1K which suggests that this transition is of nearly second order or weakly first order type.

**(D) Evidence for an AFD phase transition at low temperatures**

Since the SXRD patterns do not reveal any structural change below $T_c$~275K around which we observe an anomaly in the unit cell volume, FWHM of $(h00)_{pc}$ peaks and longitudinal elastic modulus, we collected neutron powder diffraction data on our sample in search of an AFD transition. Figs. 2(a) and (b) of the supplementary files[33] depict the synchrotron X-ray and neutron powder diffraction patterns at 100K, respectively. It is evident from this figure that new peaks with rather low intensities have appeared in the neutron powder diffraction pattern. All these peaks could be indexed as $(3/2\ 1/2\ 1/2)_{pc}$, $(3/2\ 3/2\ 1/2)_{pc}$ and $(5/2\ 1/2\ 1/2)_{pc}$ with respect to a pseudocubic cell. The presence of peaks with such odd-odd-odd half-integer indices suggests an AFD transition leading to the anti-phase rotation of the oxygen octahedra in the neighbouring unit cells[3] which are driven by phonon instabilities at the $q=(1/2\ 1/2\ 1/2)_{pc}$ point of the cubic Brilloun zone. The evolution of the intensity of the three superlattice peaks due to the AFD transition is depicted in Fig. 5 for the temperature range from 12 to 300K. It is evident from this figure that all the superlattice peaks disappear above 275K in agreement with the appearance of an anomaly in the longitudinal elastic modulus, unit cell volume and FWHM of $(h00)_{pc}$ peaks. *While we observe the $(3/2\ 1/2\ 1/2)_{pc}$, $(3/2\ 3/2\ 1/2)_{pc}$ and $(5/2\ 1/2\ 1/2)_{pc}$ superlattice peaks, the $(1/2\ 1/2\ 1/2)_{pc}$ superlattice peak is not discernible (see last column in Fig.5).*

*The tilt angle associated with an AFD transition constitutes the order parameter whose temperature dependence can be studied by investigating the temperature dependence of the integrated intensity of the superlattice peak which is proportional to the square of the tilt angle. Fig. 6 depicts the temperature dependence of the integrated intensity of the pseudocubic* (3/2 1/2 1/2) *superlattice reflection. The solid line gives the least-square fit for the power law* $I_{(3/2\ 1/2\ 1/2)pc} \sim (T_c - T)^{2\beta}$ *for which an exponent of 0.326 and* $T_c \simeq 279K$ *was determined. This value of exponent is close to 1/3. Such a value has been reported in the past by Muller and Berlinger*[36] *for I4/mcm to* $Pm\bar{3}m$ *AFD transition in* $SrTiO_3$ *and* $R\bar{3}c$ *to* $Pm\bar{3}m$ *AFD transition in* $LaAlO_3$ *using electron paramagnetic resonance (EPR) measurements. They showed that the order parameter follows second order behavior with* $\beta=1/2$ *for* $T/T_c \lesssim 0.9$ *but in the narrow* $0.9<T/T_c<1.0$ *temperature range, it deviates from* $\beta=1/2$ *and becomes* $\beta=1/3$*. They attributed it to the breakdown of the mean field behavior near* $T_c$*. However, the situation in Fig.6 is different as it does not show linear dependence above* $T/T_c \sim 0.9$ *also. We were therefore constrained to fit the power law dependence,* $I_{(3/2\ 1/2\ 1/2)pc} \sim (T_c - T)^{2\beta}$*, over the entire temperature range. This fitting gave us* $\beta=1/3$ *and* $T_c \sim 279K$*. Such a value of* $\beta=1/3$ *in recent years has been argued to be due to near tricritical behavior*[37, 38] *within the mean field approximation.*

**(E) Structure of the ground state of PSZT550**

The superlattice peaks appearing at and below $T_c \sim 275K$ in the neutron powder diffraction patterns may be due to R3c or Cc space groups[8, 11-13, 15, 18, 20-24] as discussed earlier. We analyzed the neutron diffraction data at 12K using R3c, R3c+Cm and Cc space group models. Fig. 7 compares the Rietveld fits of $(200)_{pc}$, $(220)_{pc}$ and $(111)_{pc}$ perovskite peaks for these three models. The fits for the three prominent superlattice peaks $(3/2\ 1/2\ 1/2)_{pc}$, (3/2 3/2 1/2)$_{pc}$ and (5/2 1/2 1/2)$_{pc}$ are shown in Fig. 8. It is evident from Fig. 7(a) that the R3c space group can not account for the intensity of the main perovskite reflections satisfactorily. Consideration of the coexisting Cm phase, as proposed by Frantti et al[8] and Yokota et al[20], improves the fits and $\chi^2$ drops from 8.37 to 5.50, as can be seen from Fig. 7(b). The results of refinement using the Cc space group are shown in Fig. 7(c). The Cc space group leads to a significant improvement in the fits. The $\chi^2$ (2.81) for the Cc space group is significantly lower than that for the R3c+Cm phase coexistence model (5.50). Further, the R3c or the R3c+Cm coexistence model cannot account for the (5/2 1/2 1/2)$_{pc}$ superlattice peak as shown in Fig. 8. The calculated peak position for the best Rietveld fit using R3c or R3c+Cm models is shifted towards lower angles with respect to the observed peak position. Also, the calculated profile shape cannot explain the observed profile shape of the (5/2 1/2 1/2)$_{pc}$ peak. Such a misfit in the peak position of (5/2 1/2 1/2)$_{pc}$ superlattice peak was pointed out earlier by Ranjan et al[39] for pure PZT with x=0.520 also. The Cc space group model, in contrast to R3c or R3c+Cm models, not only accounts for the (5/2 1/2 1/2)$_{pc}$ superlattice peak but also gives better fits for all other peaks, the main perovskite as well as the superlattice peaks. It is worth mentioning here that the number of refinable parameters for the R3c+Cm model (43) is higher than that (35) for the Cc model and yet it is the Cc model which gives lower $\chi^2$ and excellent overall fit. All these factors obviously favour the Cc space group. Fig. 3 of the supplementary files[33] gives the overall fit using Cc model for PSZT550. Table 1 gives the coordinates of the atoms in the Cc phase, as obtained by the refinement.

*The Cc space group permits (1/2 1/2 1/2)$_{pc}$ superlattice peak also which is otherwise extinguished for the R3c space group. Last row of Fig. 8 shows the Rietveld* *refined pattern around q=(1/2 1/2 1/2)$_{pc}$ position for all the three models. It is evident from this figure that the intensity of the calculated pattern using Cc structural model for (1/2 1/2 1/2)$_{pc}$ superlattice reflection is within the background counts of neutron diffraction pattern. The peak intensity is about 0.096% of the strongest (111)$_{pc}$ perovskite reflection in the neutron diffraction pattern. Evidently, this reflection is not discernible due to the extremely low intensity (smaller structure factor).*

*As stated earlier even the stronger superlattice reflections present in the neutron powder diffraction patterns are not discernible in the SXRD patterns. We simulated the X-ray powder diffraction pattern using the lattice parameters and coordinates obtained from Rietveld refinement of neutron diffraction data (see Fig.4 of the supplementary file[33]) for PSZT550 at 100K. Inset of this figure shows the zoomed view of superlattice (3/2 1/2 1/2)$_{pc}$ along with some part of the perovskite (111)$_{pc}$ reflection. It is evident from the inset that the intensity of this peak is within the background counts and hence cannot be discernible. The intensity of (3/2 1/2 1/2)$_{pc}$ superlattice reflection in the simulated pattern for X-rays is about 0.1% of the strongest (110)$_{pc}$ perovskite reflection which is comparable to the intensity of the (1/2 1/2 1/2)$_{pc}$ peak (0.096%) in the neutron powder diffraction pattern. The intensities of other superlattice reflections (3/2 3/2 1/2)$_{pc}$ and (5/2 1/2 1/2)$_{pc}$ are found to be still weaker. Our simulations thus suggest that the absence of all the superlattice reflections in the SXRD patterns and the (1/2 1/2 1/2)$_{pc}$ superlattice reflections in the powder neutron diffraction pattern have common origin i.e., extremely low intensity.*

**(F) Comparison with the Cc space group for PSZT530 composition**

From the large asymmetry and non-singlet nature of the $(3/2\ 1/2\ 1/2)_{pc}$ superlattice peak, we concluded in our previous work[25] on PSZT with x=0.530 (a pseudotetragonal composition) that the R3c space group for which the $(3/2\ 1/2\ 1/2)_{pc}$ reflection should be a singlet can be rejected. However, the results of Rietveld refinements for the R3c and R3c+Cm models were not given in our earlier work[25]. In order to understand the difference in the shape of the profiles and their Reitveld fits for various superlattice reflections of PSZT530 and PSZT550, we present in Fig. 9 the observed, calculated and difference profiles of PSZT530. It is evident from this figure that the calculated peak position of the $(3/2\ 1/2\ 1/2)_{pc}$ superlattice reflection is significantly shifted away from the observed peak position for the R3c and R3c+Cm structural models. In case of PSZT550, because of the pseudorhombohedral character of the monoclinic phase at room temperature and below, the $(3/2\ 1/2\ 1/2)_{pc}$ peak is nearly symmetric and therefore we do not see huge mismatch between the calculated and observed peak positions of the $(3/2\ 1/2\ 1/2)_{pc}$ superlattice peak of PSZT550, as is evident from a comparison of fits given in Fig. 9 and Fig. 8. However, the R3c as well as the R3c+Cm structural models lead to significant mismatch between the observed and calculated profiles of the $(5/2\ 1/2\ 1/2)_{pc}$ superlattice peak in both PSZT530 and PSZT550, although the mismatch is more pronounced in PSZT530 than in PSZT550. The discrepancies in the fits for the R3c and R3c+Cm models disappear for the Cc structural model suggesting clearly that both the pseudotetragonal (PSZT530) and pseudorhombohedral (PSZT550) monoclinic phases at room temperature transform, respectively, to low temperature pseudotetragonal and pseudorhombohedral superlattice phases, which have the same space group Cc.

## IV. Concluding remarks

The results presented in this and the previous work[25] conclusively establish that the ground state phase of PSZT for both the pseudotetragonal (x=0.530) and pseudorhombohedral (x=0.550) monoclinic compositions are also exhibit characteristics pseudotetragonal and pseudorhombohedral monoclinic, respectively, both having the same Cc space group. The two compositions considered by us correspond to in (x=0.530) and outside (x=0.550) the MPB region. The alternative model based on R3c+Cm or only R3c space groups leads to pronounced discrepancy between the observed and calculated profiles, especially for the superlattice peaks. However, this discrepancy decreases with increasing $Zr^{4+}$ content as one moves away from the pseudotetragonal monoclinic compositions towards the pseudorhombohedral monoclinic compositions. This is evident from a comparison of Fig. 8 with Fig. 9. It is therefore imperative to use pseudorhombohedral compositions as close to the MPB as possible to distinguish between the rhombohedral (R3c) and monoclinic (Cc) structural models for an unambiguous determination of the ground state superlattice phase of PSZT and for that reason of PZT also. The conclusions arrived at by Yokota et al[20] in favour of R3c+Cm structural model are based on the Rietveld analysis of diffraction data for compositions with $0.60 \leq x \leq 0.92$ which are far away from the MPB for which it may not be easy to distinguish between the rhombohedral and monoclinic superlattice models on the basis of the mismatch between the observed and calculated profiles of the superlattice peaks. However, the anomalous broadening of the $(h00)_{pc}$ peaks is still present for such $Zr^{4+}$ rich compositions (see Fig. 8 of Ragini et al[19] and Fig. 3 of Singh et al[14]). Pandey et al[15] have shown that this anomalous broadening can be explained better using Cc space group than the R3c space group.

The main objection against the Cc space group is the non-observation of a peak at q= $(1/2\ 1/2\ 1/2)_{pc}$ position which is extinguished for the R3c space group but is permitted by the Cc space group. It may be noted that the calculated intensity of the $(1/2\ 1/2\ 1/2)_{pc}$ superlattice peak *(~0.096% of strongest intensity $(111)_{pc}$ perovskite reflection for PSZT550 in neutron diffraction pattern)* is rather too low to be observable above the background level. Further, the non-observation of this extremely low intensity peak in neutron powder diffraction data is as enigmatic as the non observation of the rather intense superlattice reflections (in neutron diffraction patterns) at q= $(3/2\ 1/2\ 1/2)_{pc}$ , $(3/2\ 3/2\ 1/2)_{pc}$ and $(5/2\ 1/2\ 1/2)_{pc}$ positions in synchrotron X-ray powder diffraction patterns. We showed that the calculated intensities of these non-observable superlattice peaks in the SXRD patterns is $\leq 0.1\%$ which is comparable to the non-observable $(1/2\ 1/2\ 1/2)_{pc}$ superlattice peak in the neutron powder patterns. It was because of the non-observation of the intense superlattice peaks in the SXRD patterns that Noheda et al[16] missed the superlattice phase, discovered later on by Ragini et al[4] and Ranjan et al[5], in their first set of experiments at 20K and erroneously concluded Cm space group for the ground state of PZT in the MPB region.

Based on the analysis of the observed superlattice peaks, we conclude that the structural models for the ground state phase of PZT based on R3c space group with or without the consideration of the coexistence of the room temperature monoclinic phase in the Cm space group can be rejected and that the space group of the ground state of PZT corresponding to the compositions in the MPB and on the $Zr^{4+}$ rich side of the MPB is Cc.

**Acknowledgements:**

R. S. Solanki acknowledges financial support from Council of Scientific and Industrial Research (CSIR), India in the form of a Junior Research Fellowship. D. Pandey and Y. Kuroiwa acknowledge financial support from Department of Science and Technology (DST), Govt. of India and Japan Society for the Promotion of Science (JSPS) of Japan under the Indo-Japan Science Collaboration Program. The synchrotron radiation experiments were performed at the BL02B2 beamline of Spring-8 with the approval of Japan Synchrotron Radiation Research Institute (Proposal Nos. 2011A1324 and 2011A0084). T. Suzuki and I. Ishii thank the support (the Grant-in-Aid for Scientific Research on Innovative Areas "Heavy Electrons" (No.20102005)) from the Ministry of Education, Culture, Sports, Science and Technology of Japan.

## References:

1. B. Jaffe, W. R. Cook, and H. Jaffe, *Piezoelectric Ceramics* (Academic Press, London, 1971), p. 135.

2. C. Michel, J.-M. Moreau, G. D. Achenbach, R. Gerson, and W. J. James, Solid State Commun. **7**, 865 (1969).

3. A. M. Glazer, Acta Cryst. **B28**, 3384 (1972); A. M. Glazer, Acta Cryst. **A31**, 756 (1975).

4. Ragini, S. K. Mishra, Dhananjai Pandey, Herman Lemmens and G. Van Tendeloo, Phys. Rev. B **64**, 054101(2001).

5. Rajeev Ranjan, Ragini, S. K. Mishra, Dhananjai Pandey and Brendan J. Kennedy, Phys. Rev. B **65**, 060102(R) (2002).

6. M. Fornari and D. J. Singh, Phys. Rev. B **63**, 092101 (2001).

7. B. Noheda, L. Wu and Y. Zhu, Phys. Rev. B **66**, 060103(R) (2002).

8. J. Frantti, S. Ivanov, S. Eriksson, H. Rundlof, V. Lantto, J. Lappalainen, and M. Kakihana, Phys. Rev. B **66**, 064108 (2002); J. Frantti, J. Phys. Chem. B **112**, 6521 (2008).

9. A. M. Glazer, P. A. Thomas, K. Z. Baba-Kishi, G. K. H. Pang, and C. W. Tai, Phys. Rev. B, **70**, 184123 (2004).

10. Igor A. Kornev, L. Bellaiche, P.-E. Janolin, B. Dkhil, and E. Suard, Phys. Rev. Lett. **97,** 157601(2006).

11. David I. Woodward, Jesper Knudsen, and Ian M. Reaney, Phys. Rev. B, **72**, 104110 (2005).

12. D. E. Cox, B. Noheda, and G. Shirane, Phys. Rev. B **71**, 134110 (2005).

13. J. Rouquette, J. Haines, V. Bornand, M. Pintard, Ph. Papet, W. G. Marshall and S. Hull, Phys. Rev. B, **71**, 024112 (2005).

14. Akhilesh Kumar Singh, Dhananjai Pandey, Songhak Yoon, Sunggi Baik and Namsoo Shin, Appl. Phys. Lett., **91**, 192904(2007).

15. D. Pandey, A. K. Singh, and S. Baik, Acta Cryst. A **64**, 192(2008).

16. B. Noheda, D. E. Cox, G. Shirane, J. A. Gonzalo, L. E. Cross and S-E. Park, Appl. Phys. Lett. , **74**, 2059(1999).

17. B. Noheda, J. A. Gonzalo, L. E. Cross, R. Guo, S.-E. Park, D. E. Cox and G. Shirane, Phys. Rev. B, **61**, 8687(2000).

18. D. M. Hatch, H. T. Stokes, Rajeev Ranjan, Ragini, S. K. Mishra, Dhananjai Pandey and Brendan J. Kennedy, Phys. Rev. B **65**, 212101(2002).

19. Ragini, Rajeev Ranjan, S. K. Mishra, and Dhananjai Pandey, J. Appl. Phys, **92**, 3266(2002).

20. H. Yokota, N. Zhang, A. E. Taylor, P. A. Thomas and A. M. Glazer, Phys. Rev. B **80**, 104109(2009).

21. N. Zhang, H. Yokota, A. M. Glazer and P. A. Thomas, Acta Cryst., **B67**,386(2011).

22. D. Phelan, X. Long, Y. Xie, Z.-G. Ye, A. M. Glazer, H. Yokota, P. A. Thomas, and P. M. Gehring, Phys. Rev. Lett. **105**, 207601(2010).

23. S. Gorfman, D. S. Keeble, A. M. Glazer, X. Long, Y. Xie, Z.-G. Ye, S. Collins, and P. A. Thomas, Phys. Rev. B **84**, 020102(R) (2011).

24. Guillaume Fraysse, Julien Haines, Véronique Bornand, Jérôme Rouquette, Marie Pintard, Philippe Papet and Steve Hull, Phys. Rev. B **77**, 064109(2008).

25. Ravindra Singh Solanki, Akhilesh Kumar Singh, S. K. Mishra, Shane J. Kennedy, Takashi Suzuki, Yoshihiro Kuroiwa, Chikako Moriyoshi, and Dhananjai Pandey, Phys. Rev. B, **84**, 144116 (2011).

26. A. P. Singh, S. K. Mishra, D. Pandey, Ch. D. Prasad and R. Lal, J. Mater. Sci. **28**, 5050 (1993).

27. E. Nishibori, M. Takata, K. Kato, M. Sakata, Y. Kubota, S. Aoyagi, Y. Kuroiwa, M. Yamakata, and N. Ikeda, Nucl. Instrum. Methods Phys. Res. A **467-468**, 1045 (2001).

28. M. Hoelzel, A. Senyshyn, R. Gilles, H. Boysen, and H. Fuess, Neutron News **18** (4), 23 (2007).

29. T. Goto, T. Suzuki, A. Tamaki, Y. Ohe, S. Nakamura, and T. Fujimura, *The Bulletin of the Research Institute for Scientific Measurement* (Tohoku University, Sendai, Japan, 1989), Vol. 38, p. 65.

30. J. Rodriguez-Carvajal Laboratory, FULLPROF, a Rietveld and pattern matching and analysis program version July 2011, Laboratoire Leon Brillouin, CEA-CNRS, France [http://www.ill.eu/sites/fullprof/].

31. Akhilesh Kumar Singh, Oksana Zaharkoand Dhananjai Pandey Phys. Rev. B, **68**, 172103(2003).

32. Peter W. Stephens, J. Appl. Cryst. **32**, 281(1999).

33. See supplementary file material at [URL will be inserted by publisher] for a comparison of synchrotron X-ray and neutron powder diffraction pattern at 300 and 100K. It also contains Rietveld fit of neutron powder diffraction pattern at 12K, calculated/simulated pattern at 100K and refined structural parameters at 300 and 100K obtained after Rietveld analysis of SXRD data.

*34. R. G. Burkovsky, Yu. A. Bronwald, A. V. Filimonov, A. I. Rudskoy, D. Chernyshov, A. Bosak, J. Hlinka, X. Long, Z.-G. Ye, and S. B. Vakhrushev, Phys. Rev. Lett.* ***109,*** *097603 (2012).*

35. A. P. Wilkinson, J. Xu, S. Pattanaik and S. J. L. Billinge, Chem. Mater. **10**, 3611 (1998).

*36. K. A. Muller and W. Berlinger, Phys. Rev. Lett.* ***26****, 13 (1971).*

*37. E. K. H. Salje, J. Mc Jimenz Gallardo, F. J. Romero, and J. del Cerro, J. Phys.: Condens. Matter* ***10****, 5535 (1998).*

*38. Sanjay Kumar Mishra and Dhananjai Pandey, Phys. Rev. B* ***79****, 174111 (2009).*

39. Rajeev Ranjan, Akhilesh Kumar Singh, Ragini, and Dhananjai Pandey, Phys. Rev. B **71**, 092101(2005).

## Figure Captions:

**Figure 1.** The evolution of synchrotron X-ray powder diffraction profiles of the $(111)_{pc}$, $(200)_{pc}$ and $(220)_{pc}$ reflections of PSZT550 with temperature.

**Figure 2.** (Color online) Observed (red dots), calculated (black continuous line) and difference (bottom line) profiles at 300K using R3m and Cm space groups obtained after Rietveld refinement. The vertical tick marks above the difference profiles give the positions of the Bragg reflections. Insets show the magnified view of Rietveld fits for selected $(111)_{pc}$, $(200)_{pc}$ and $(220)_{pc}$ reflections.

**Figure 3**. (Color online) Observed (red dots), calculated (black continuous line) and difference (bottom line) profiles at 100K using R3m and Cm space groups obtained after Rietveld refinement. The vertical tick marks above the difference profiles give the positions of the Bragg reflections. Insets show the magnified view of Rietveld fits for selected $(111)_{pc}$, $(200)_{pc}$ and $(220)_{pc}$ reflections.

**Figure 4.** Temperature variation of (a) unit cell volume, (b) FWHM of the $(200)_{pc}$ peak and (c) longitudinal elastic modulus ($C_L$) of PSZT550. Inset of (c) shows a magnified view of the phase transition region during heating and cooling cycles.

**Figure 5.** (Color online) The evolution of neutron powder diffraction profiles of the $(3/2\ 1/2\ 1/2)_{pc}$, $(3/2\ 3/2\ 1/2)_{pc}$ and $(5/2\ 1/2\ 1/2)_{pc}$ superlattice reflections of PSZT550 as a function of temperature. *(Intensities are normalized for comparison purposes and are plotted on the same scales.)*

***Figure 6.*** *Variation in the integrated intensity of the $(3/2\ 1/2\ 1/2)_{pc}$ superlattice peak with temperature. Dots represent experimental values, while the continuous line corresponds to the least-squares fit for $I_{(3/2\ 1/2\ 1/2)pc} \sim (T_c-T)^{2\beta}$.*

**Figure 7.** (Color online) Observed (red dots), calculated (black continuous line) and difference (bottom line) plots of the three perovskite peaks $(111)_{pc}$, $(200)_{pc}$, and $(220)_{pc}$ of PSZT550 at 12K, as obtained by Rietveld refinement using (i)R3c, (ii) R3c+Cm and (iii) Cc structural models.

**Figure 8.** (Color online) Observed (red dots), calculated (black continuous line) and difference (green bottom line) plots of the three superlattice peaks $(3/2\ 1/2\ 1/2)_{pc}$, $(3/2\ 3/2\ 1/2)_{pc}$ and $(5/2\ 1/2\ 1/2)_{pc}$ of PSZT550 at 12K, as obtained by Rietveld refinement using (i)R3c, (ii) R3c+Cm and (iii) Cc structural models.

**Figure 9.** (Color online) Observed (red dots), calculated (black continuous line) and difference (bottom line) plots of three superlattice reflections $(3/2\ 1/2\ 1/2)_{pc}$, $(3/2\ 3/2\ 1/2)_{pc}$ and $(5/2\ 1/2\ 1/2)_{pc}$ obtained by Rietveld refinement of PSZT530 using (i) R3c, (ii) R3c+Cm and (iii) Cc structural models. The vertical tick marks above the difference profiles give the positions of the Bragg reflections.

## Table Captions:

**Table1.** Refined structural parameters and agreement factors for PSZT550 using neutron powder diffraction data at 12K with Cc space group.

**Table1.**

| T=12K, Cc space group | | | | |
|---|---|---|---|---|
| a= 10.0316(2)Å, b= 5.7462 (2)Å, c= 5.7748 (2)Å, β= 125.637(2) | | | | |
| $\chi^2$= 2.32, $R_{wp}$ = 4.76 | | | | |
| Atoms | x | y | z | B(Å$^2$) |
| Pb/Sr | 0 | 0.75 | 0 | $\beta_{11}$=0.001(1), $\beta_{22}$=0.00, $\beta_{33}$ =0.007(2), $\beta_{12}$=0.005(1), $\beta_{13}$=0.0058(8), $\beta_{23}$=0.013(1) |
| Ti/Zr | 0.219(5) | 0.253(6) | 0.247(3) | 0.02(3) |
| O1 | -0.029(1) | 0.270(2) | 0.370(3) | 0.81(3) |
| O2 | 0.219(1) | 0.520(2) | 0.0373(3) | 0.19(1) |
| O3 | 0.205(1) | -0.007(2) | 0.500(1) | 0.33(2) |

**Figures**

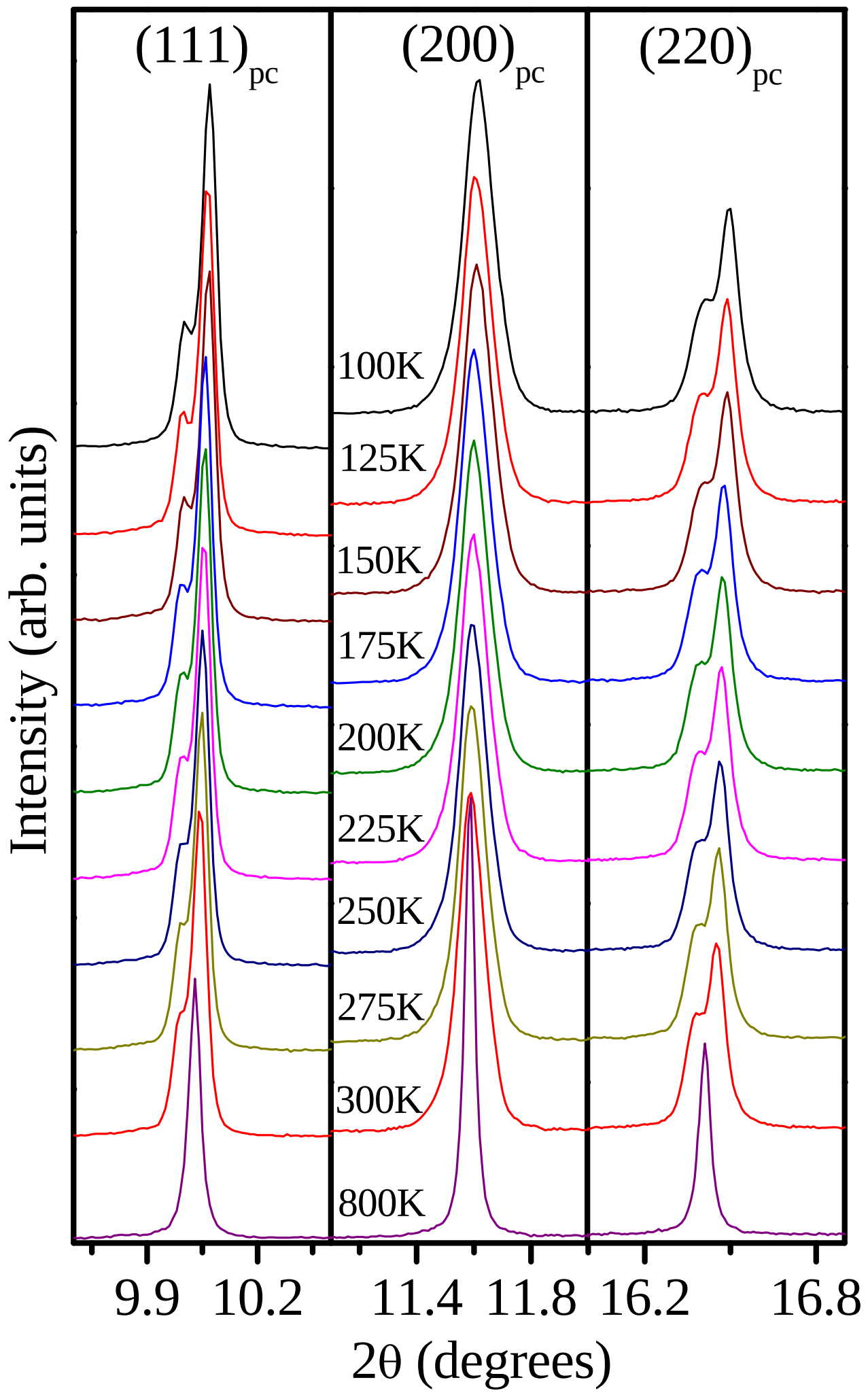

**Fig.1**

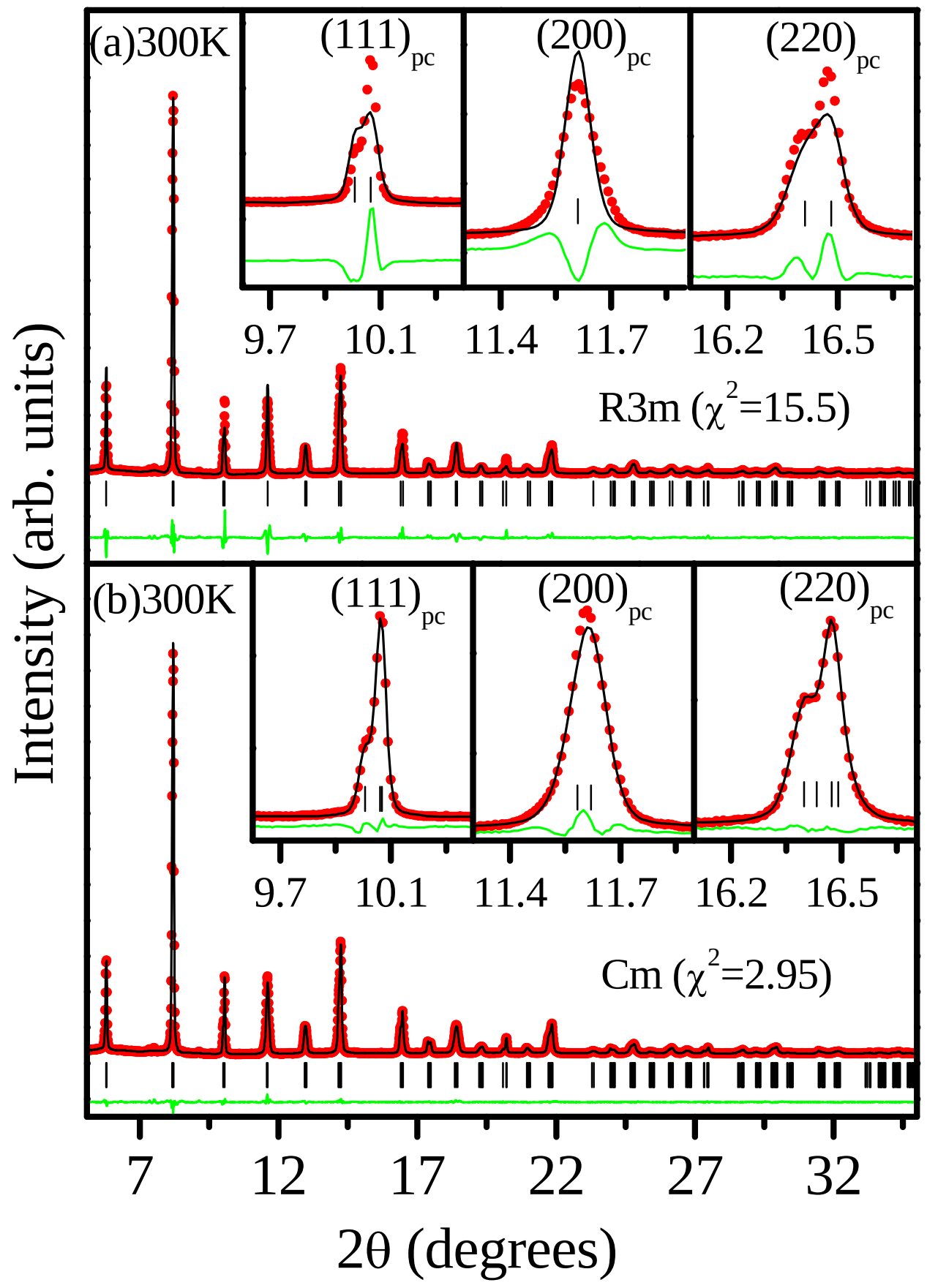

**Fig.2**

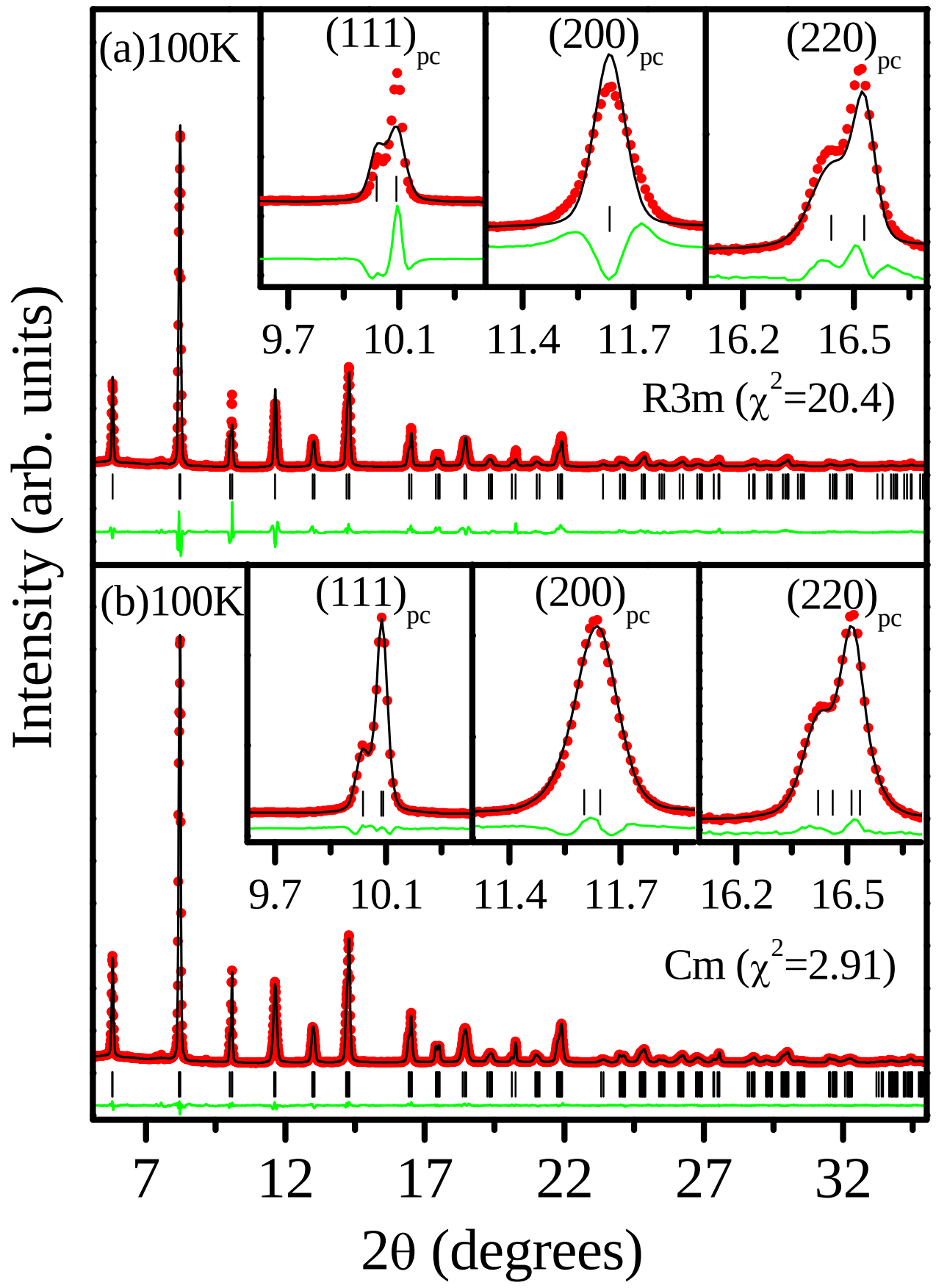

**Fig.3**

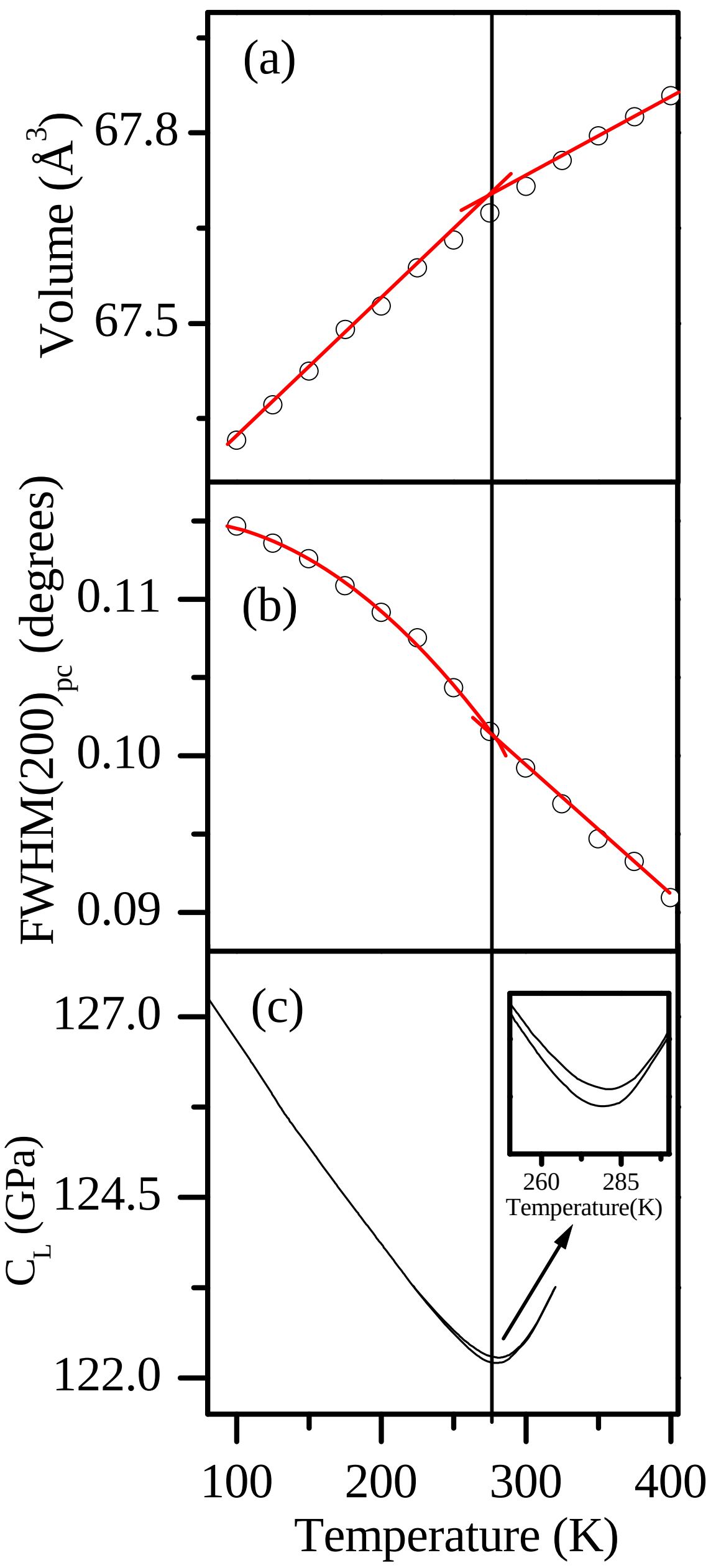

**Fig. 4**

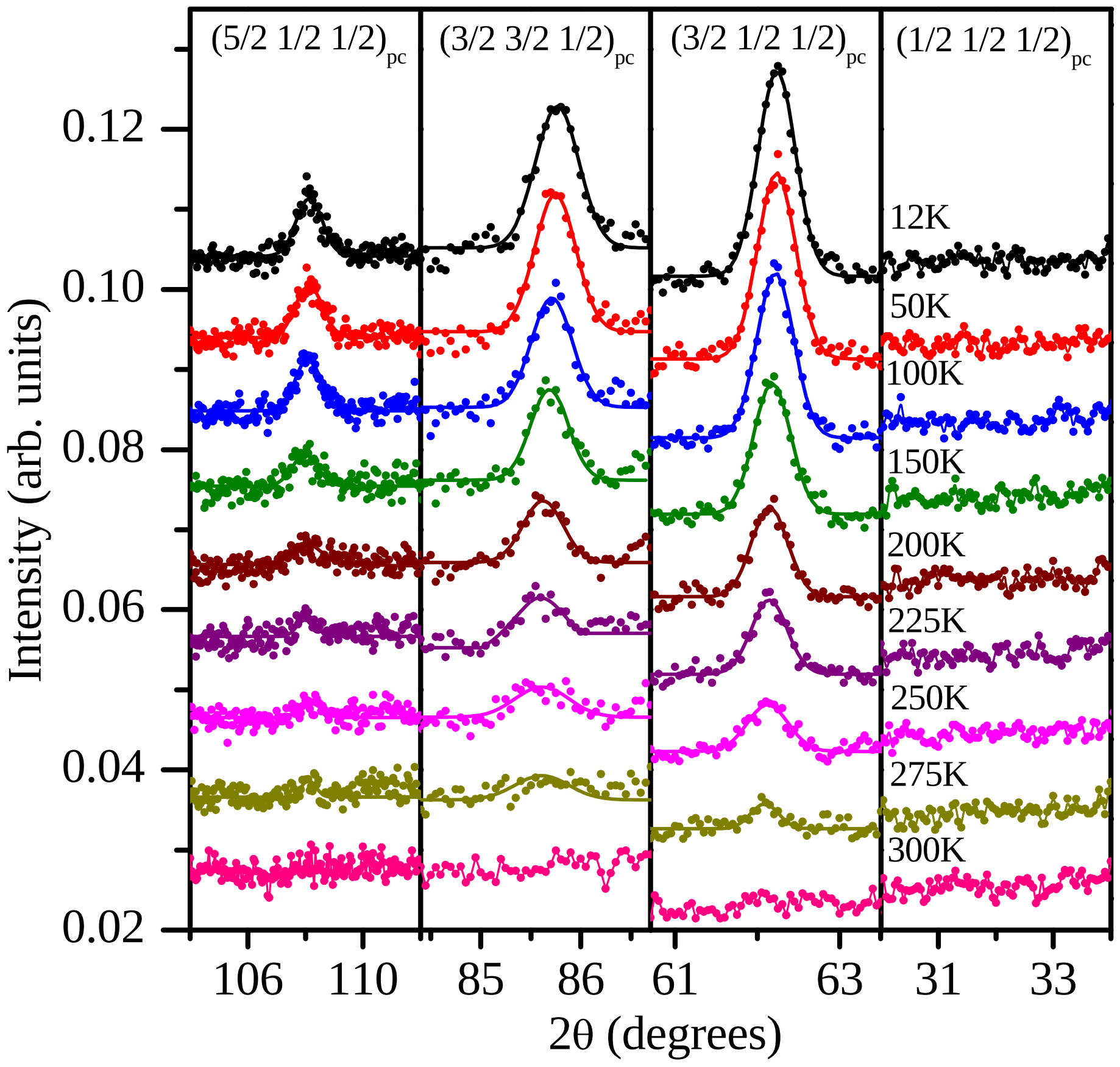

**Fig. 5**

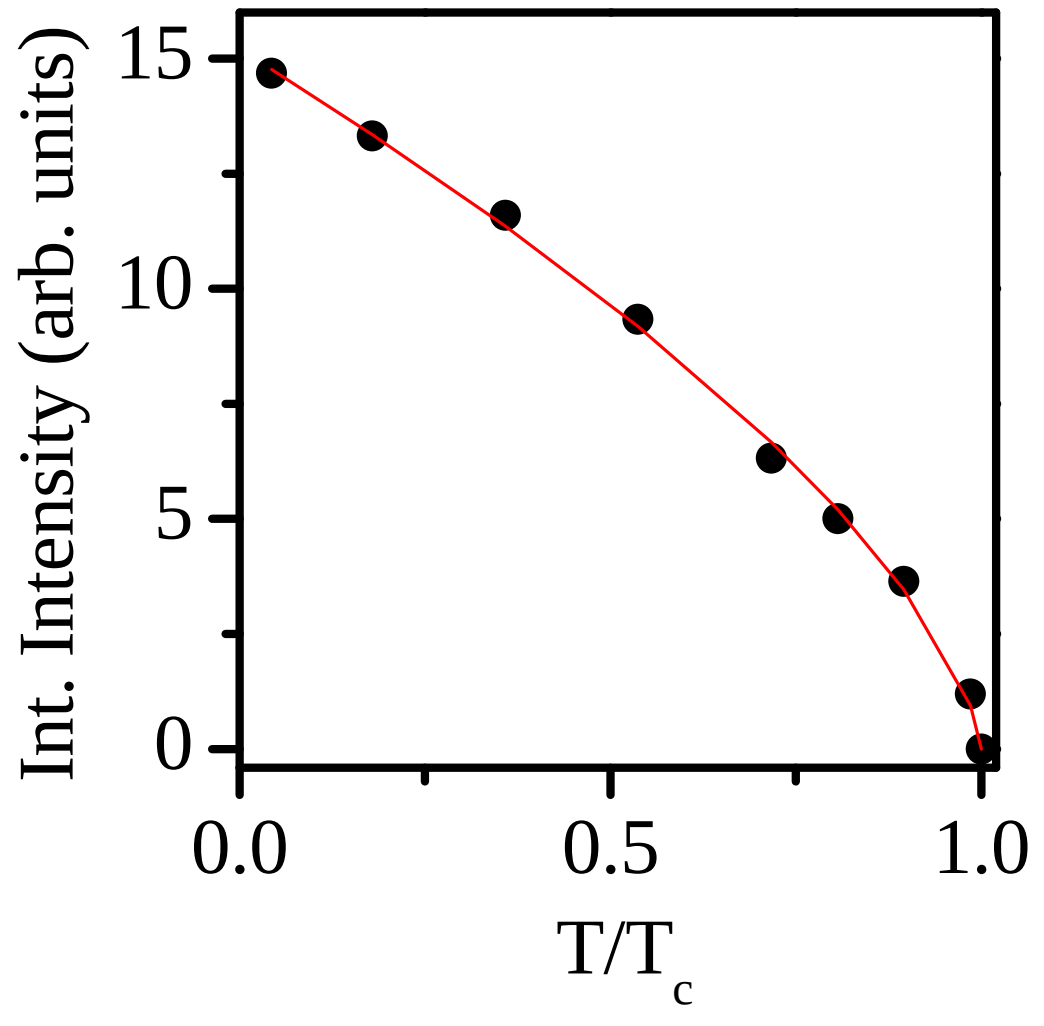

**Fig. 6**

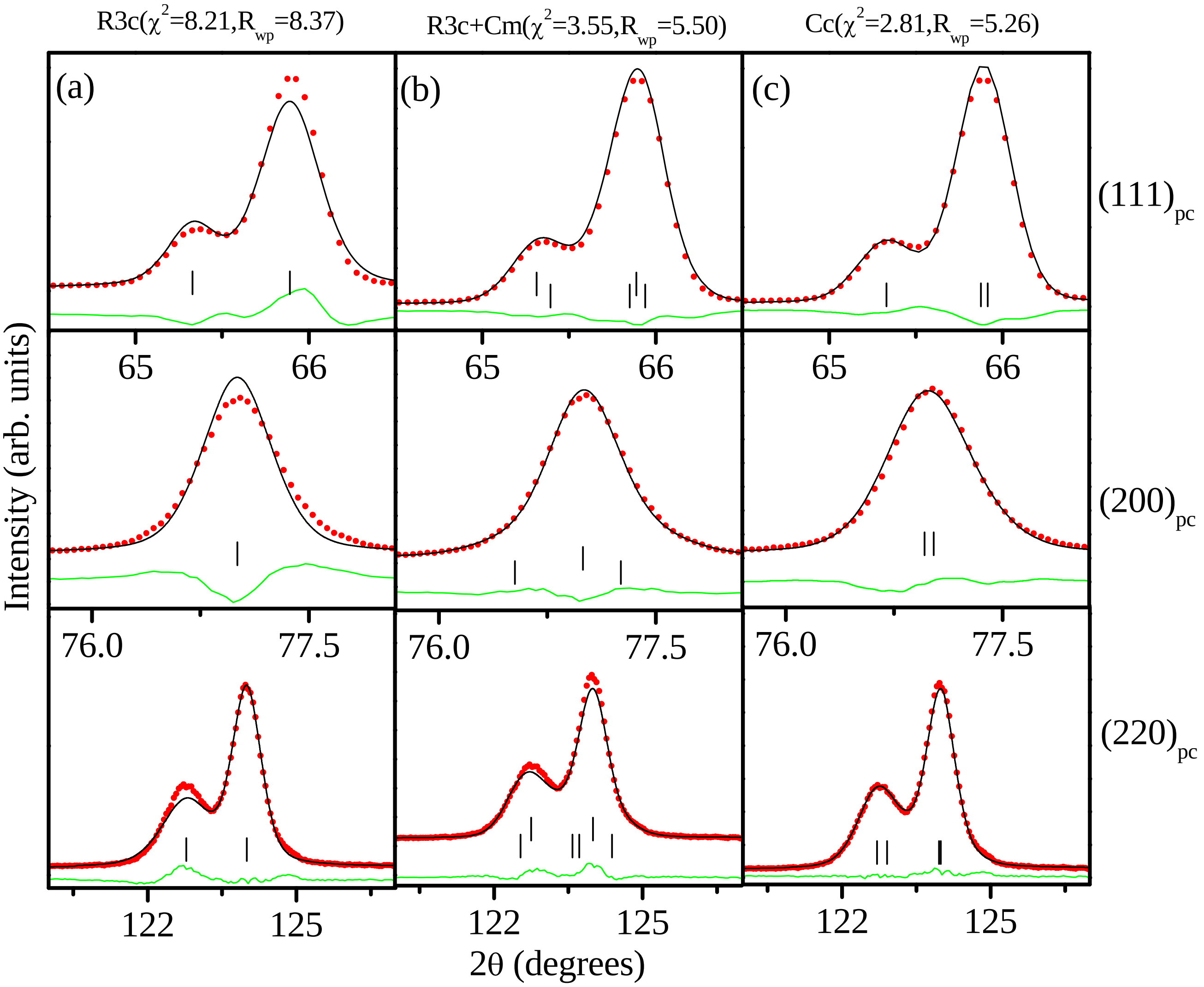

**Fig. 7**

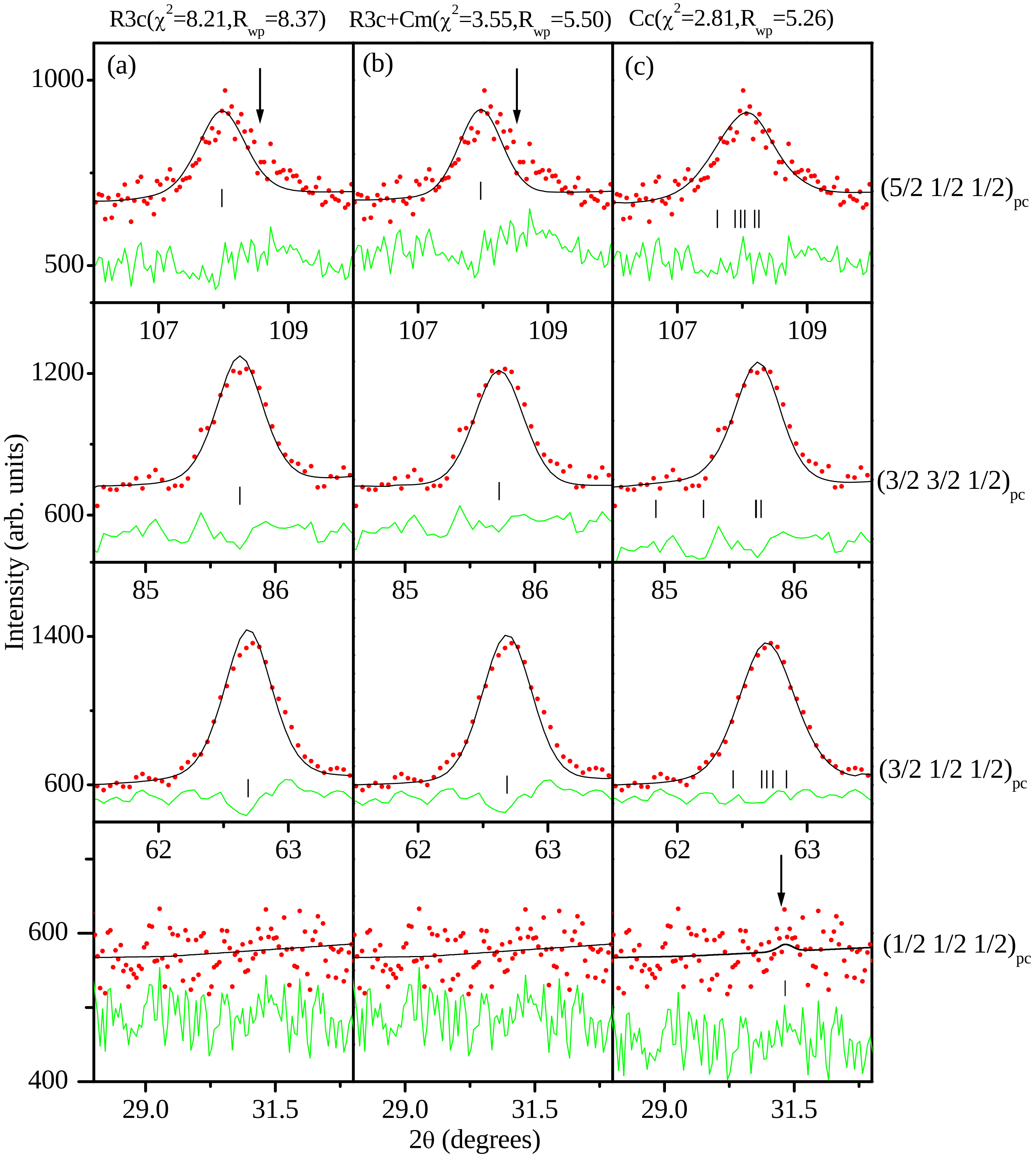

**Fig. 8**

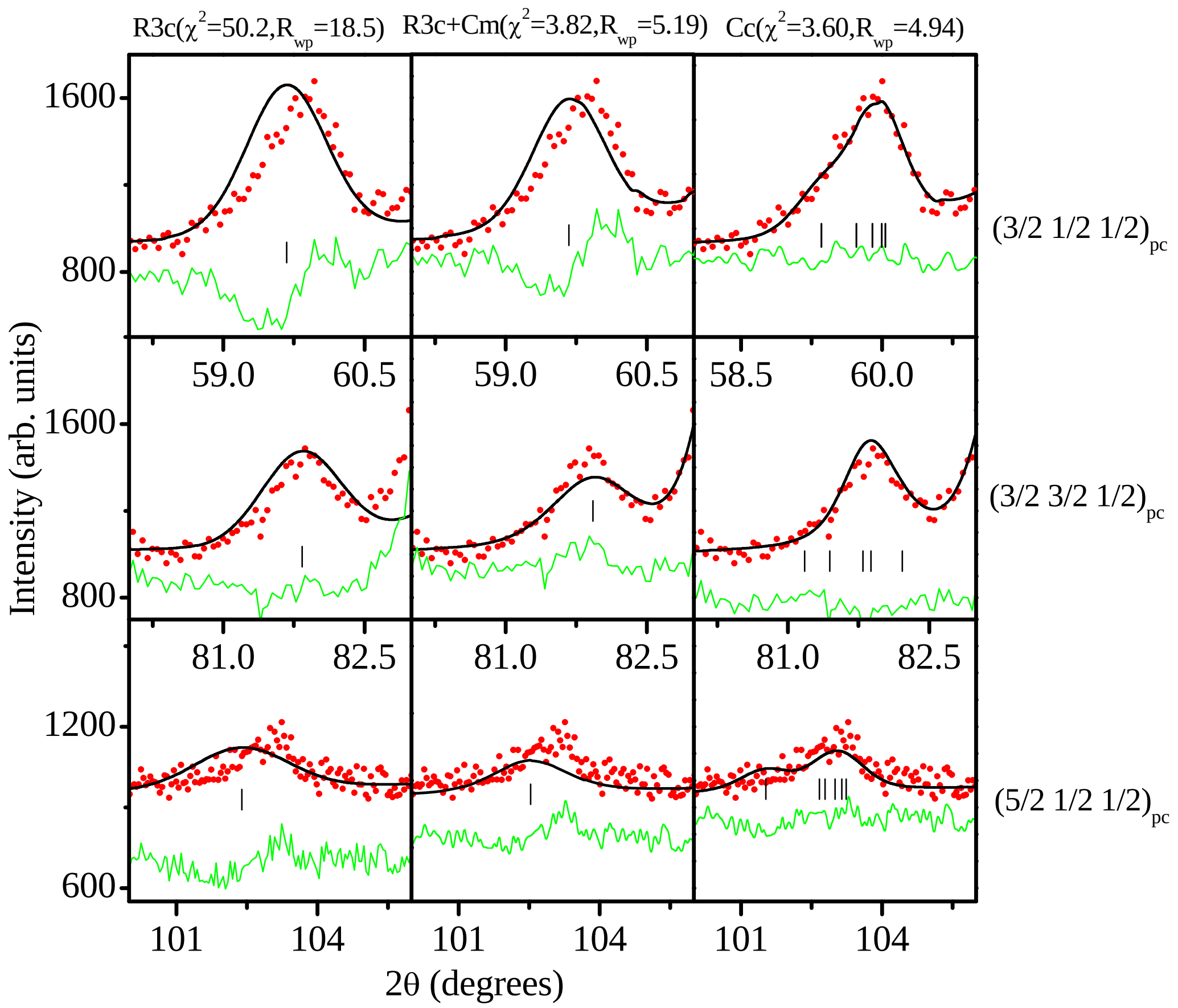

**Fig. 9**

**SUPPLEMENTARY FILE**

**Fig.1.** Powder diffraction patterns of PSZT550 at 300K: (a) high resolution synchrotron data at λ=0.412 Å and (b) high resolution neutron powder diffraction data recorded at λ= 2.5363 Å. No superlattice reflection is visible in both the diffraction patterns.

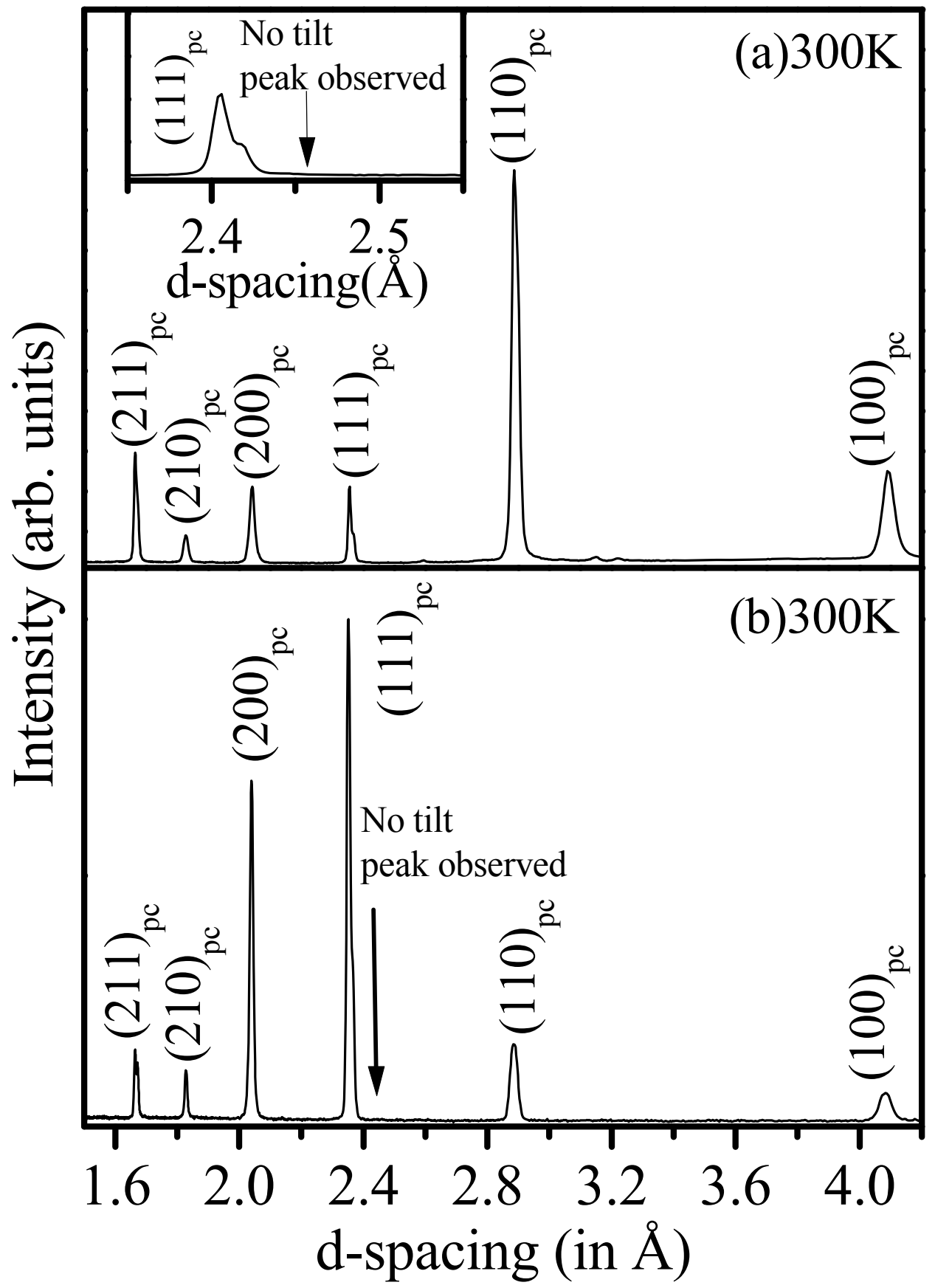

**Fig.2.** Powder diffraction patterns of PSZT550 at 100K: (a) high resolution synchrotron data at λ=0.412 Å and (b) high resolution neutron powder diffraction data recorded at λ= 2.5363 Å. The superlattice reflections with half integer indices are clearly seen in (b).

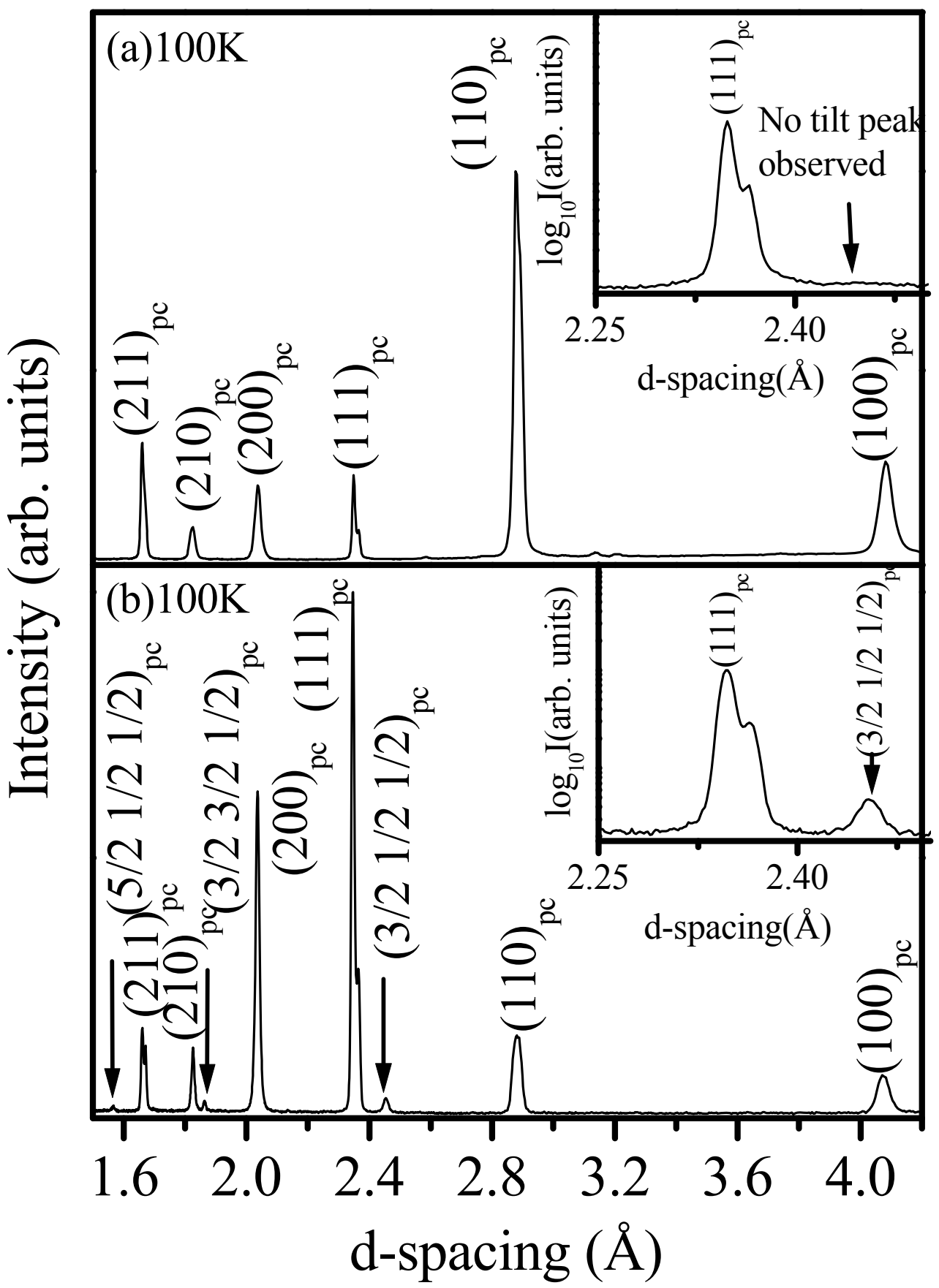

**Fig.3.** (Color online) The overall fit between the observed (red dots), calculated (black continuous line) and difference (bottom line) plots obtained by Rietveld refinement using Cc space group at 12K using neutron powder diffraction data. The vertical tick marks above the difference profiles give the positions of the Bragg reflections.

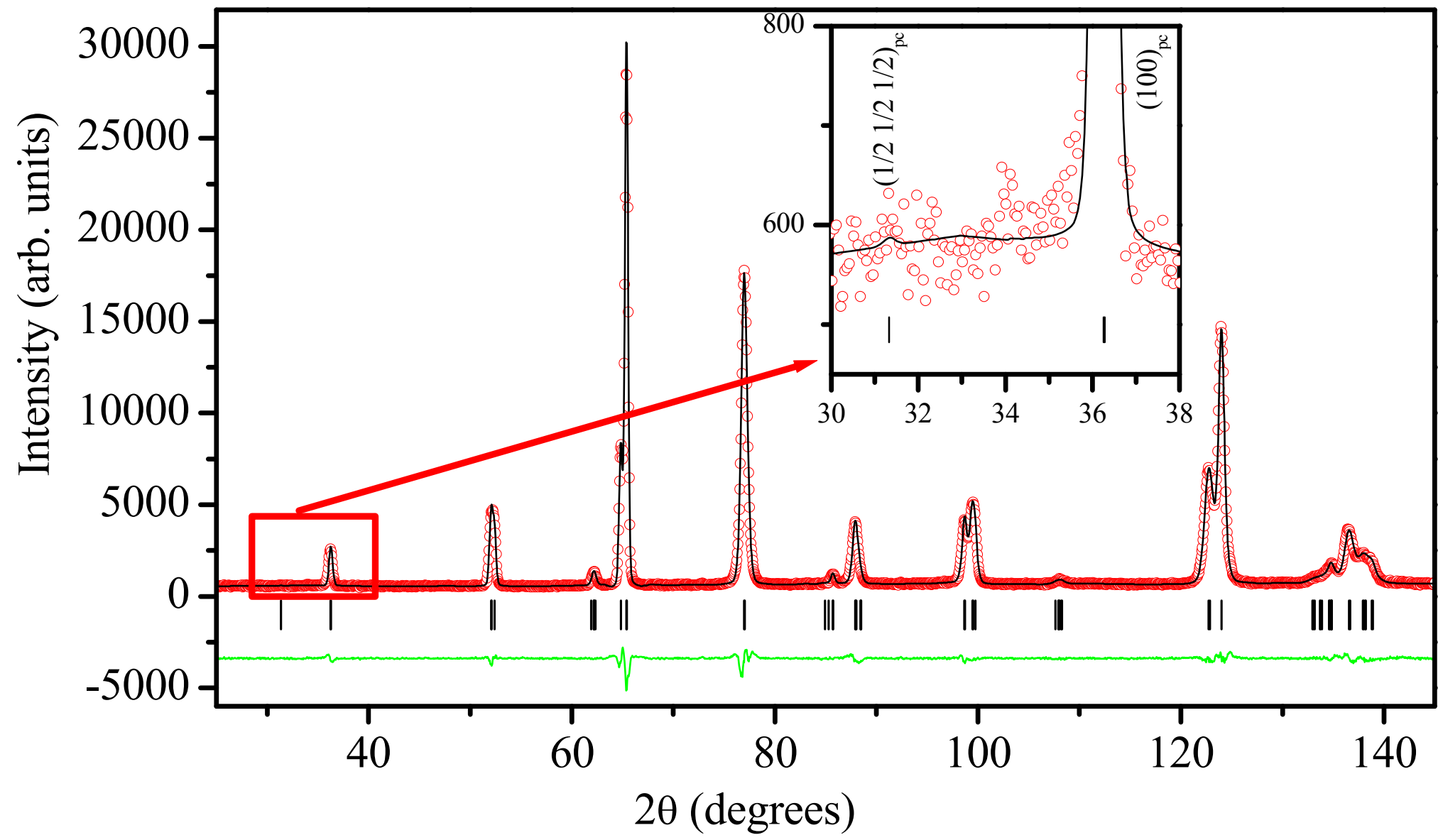

**Fig.4.** (Color online) Observed (red dots), simulated (black continuous line) and difference (bottom line) X-ray diffraction pattern of PSZT550 at 100K with Cc space group using structural parameters obtained from neutron data. The vertical tick marks above the difference profiles give the positions of the Bragg reflections.

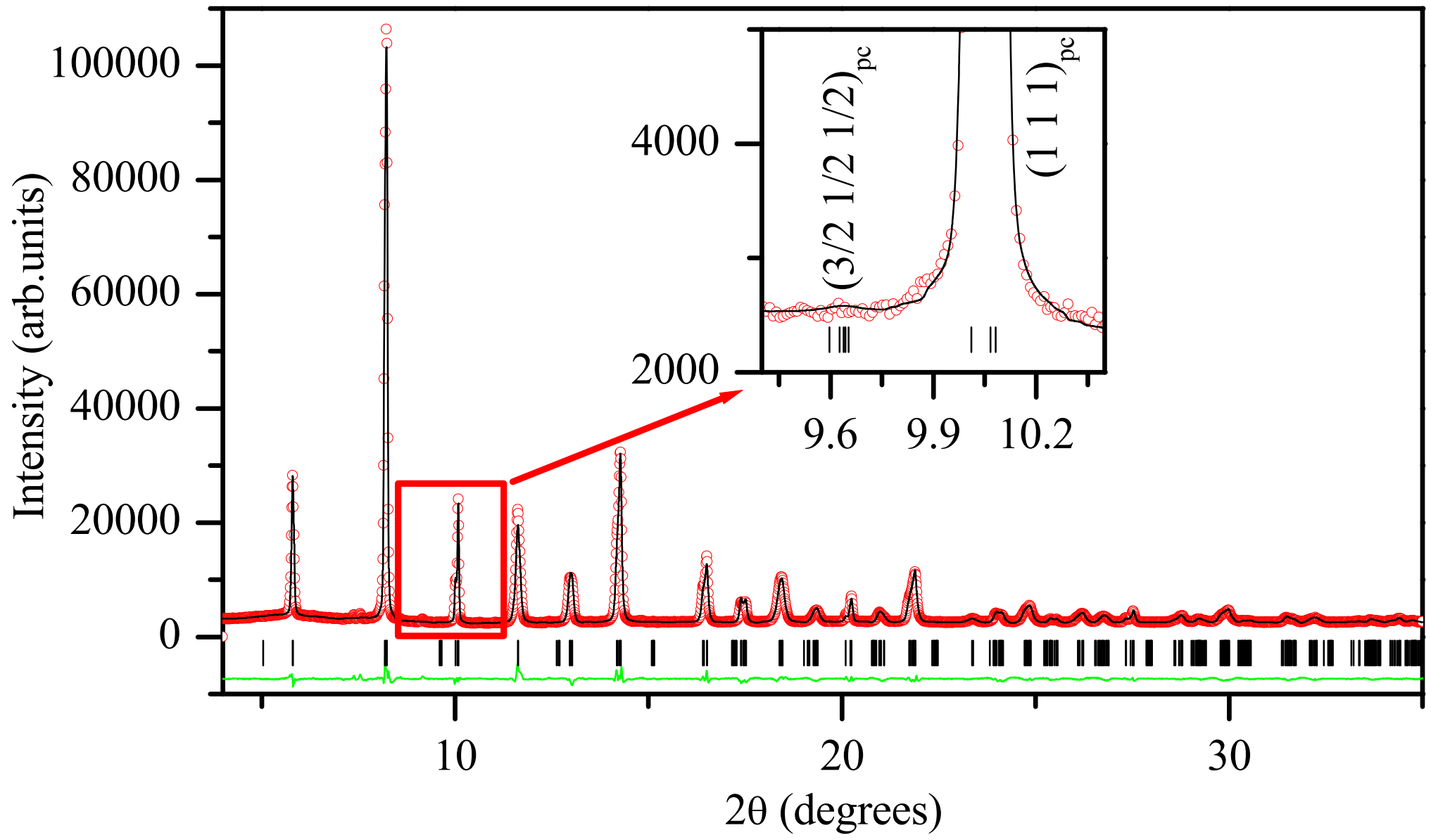

**Table1.** Refined structural parameters and agreement factors for PSZT550 using SXRD data at 300K and 100K with Cm space group.

| T=300K, Cm space group<br>a=5.7669(3) Å, b=5.7470(2) Å,<br>c=4.08651(2) Å, β= 90.367(2),<br>$\chi^2$= 2.95, $R_{wp}$ = 2.73 | | | | | T=100K, Cm space group<br>a= 5.7569(2) Å, b= 5.7310(2) Å,<br>c= 4.0794(2) Å, β= 90.426(2),<br>$\chi^2$= 2.91, $R_{wp}$ = 2.74 | | | |
|---|---|---|---|---|---|---|---|---|
| Atoms | x | y | z | B(Å$^2$) | x | y | z | B(Å$^2$) |
| Pb/Sr | 0.00 | 0.00 | 0.00 | $\beta_{11}$=0.015(1), $\beta_{22}$=0.018(1), $\beta_{33}$ =0.017(2), $\beta_{13}$=0.015(8) | 0.00 | 0.00 | 0.00 | $\beta_{11}$=0.008(1), $\beta_{22}$=0.008(2), $\beta_{33}$ =0.006(2), $\beta_{13}$=0.007(8) |
| Ti/Zr | 0.520(1) | 0.000 | 0.458(1) | 0.002 | 0.524(1) | 0.000 | 0.457(2) | 0.002 |
| O1 | 0.518(6) | 0.000 | -0.059(5) | 1.55(2) | 0.532(5) | 0.000 | -0.066(5) | 1.02(3) |
| O2 | 0.286(2) | 0.252(3) | 0.0394(3) | 0.09(2) | 0.298(3) | 0.249(2) | 0.381(2) | 0.07(1) |